# Determination of the modes in two types of closed circuits with quantum tunneling

Mark J. Hagmann

*Department of Electrical and Computer Engineering, University of Utah, 50 S. Central Campus Dr #2110, Salt Lake City, Utah 84112, USA*
*Corresponding author: Mark J. Hagmann, email: Mark.Hagmann@utah.edu*

**ABSTRACT**

Others have solved the Schrödinger equation for a one-dimensional model having a square potential barrier in free-space by requiring an incident and a reflected wave in the semi-infinite pre-barrier region, two opposing waves in the square barrier, and a transmitted wave in the semi-infinite post-barrier region. Now we model a pre-barrier region of finite length that is shunted by the barrier to form a closed circuit. We use the boundary condition that the wavefunction and its derivative are continuous at the both ends of this model to obtain a homogeneous matrix equation. Thus, the determinant must be zero for a non-trivial solution. All but one of the following four parameters are specified and the remaining one is varied to bring the determinant to zero: (1) the electron energy, (2) the barrier length, (3) the barrier height, and (4) the pre-barrier length. The solutions with a square barrier are sets of non-intersecting S-shaped lines in this four-parameter space. The solutions with a triangular barrier have the product of the propagation constant and the length of the pre-barrier region as integer multiples of two-pi radians. Only static solutions are considered, but this method could be applied to time-dependent cases under quasistatic conditions. Suggestions are given for the design and testing of prototypes.

## I. INTRODUCTION

Others have used one-dimensional models to simulate the interaction of a charged particle with a square potential barrier [1] as well as a triangular barrier [2] by specifying the potential to an infinite distance beyond both ends of the barrier. When there is an incident wave at one end of the barrier the solution of the Schrödinger equation requires a reflected wave on that side, a transmitted wave on the other side, and two waves with opposite directions within the barrier. Now we present two models for closed tunneling circuit; the first with a square barrier, and the second with a triangular barrier.

## II. CLOSED-CIRCUIT WITH A SQUARE BARRIER

Suppose that two cylindrical electrically-conductive tubes, that are held at potentials of zero and $V_{II}$, are bent and attached at their ends to form a circular loop. We consider the possibility that if electrons with energy E are injected into the tubes, they could be constrained to move on circular paths within these tubes by using a magnetic field or other means. We also make the approximation of neglecting the effects of the magnetic field when applying the Schrödinger equation. Thus, in Fig. 1 we approximate this circuit where Region I is the part where the potential is zero and Region II is the part where the potential is $V_{II}$. Here "x" denotes the coordinate at points along the closed circuit instead of its typical use as a Cartesian coordinate. Also, note that a is less than zero and b is greater than zero.

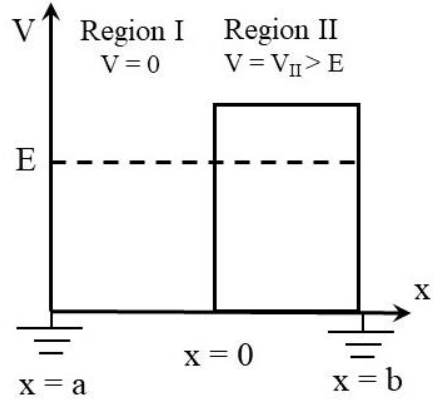

Fig. 1. Potential in Regions I and II with a square barrier.

We follow the convention that the wavefunction has a time-dependence of $\exp(-iEt/\hbar)$ where E is the energy of the particle. Thus, the solution in the two regions shown in Fig. 1 is given by Eqs. 1 and 2, where k and β are defined in Eqs. 3 and 4. Here the energy has units of electron volts and the potential is in volts so the magnitude of the electron charge "e" is included in Eqs. 3 and 4. The derivative of the wavefunction in the two regions is shown in Eqs. 5 and 6.

$$\psi_I = A\cos(kx) + B\sin(kx) \tag{1}$$

$$\psi_{II} = Ce^{\beta x} + De^{-\beta x} \tag{2}$$

$$k \equiv \frac{\sqrt{2emE}}{\hbar} \tag{3}$$

$$\beta \equiv \frac{\sqrt{2me(V_{II} - E)}}{\hbar} \tag{4}$$

$$\frac{d\psi_I}{dx} = -kA\sin(kx) + kB\cos(kx) \tag{5}$$

$$\frac{d\psi_{II}}{dx} = \beta Ce^{\beta x} - \beta De^{-\beta x} \tag{6}$$

Applying the boundary conditions so that the wavefunction and its derivative are continuous for x equal to zero gives Eqs. 7 and 8.

$$A - C - D = 0 \tag{7}$$

$$kB - \beta C + \beta D = 0 \tag{8}$$

Requiring continuity of the wave function and its derivative at the boundary where x equals a in region I to that where x equals b in region II gives Eqs. 9 and 10.

$$A\cos(ka) + B\sin(ka) - Ce^{\beta b} - De^{-\beta b} = 0 \tag{9}$$

$$kA\sin(ka) - kB\cos(ka) + \beta Ce^{\beta b} - \beta De^{-\beta b} = 0 \tag{10}$$

The system of Eqs, 7, 8, 9, and 10 in the four coefficients A, B, C, and D may be written in matrix form as shown in Eq. 11.

$$\begin{bmatrix} +1 & 0 & -1 & -1 \\ 0 & +k & -\beta & +\beta \\ +\cos(ka) & +\sin(ka) & -e^{\beta b} & -e^{-\beta b} \\ +k\sin(ka) & -k\cos(ka) & +\beta e^{\beta b} & -\beta e^{-\beta b} \end{bmatrix} \begin{bmatrix} A \\ B \\ C \\ D \end{bmatrix} = \begin{bmatrix} 0 \\ 0 \\ 0 \\ 0 \end{bmatrix} \quad (11)$$

However, this is a homogeneous system of equations so to have a non-trivial solution for the four coefficients the determinant must be zero, as shown in Eq. 12.

$$\begin{vmatrix} +1 & 0 & -1 & -1 \\ 0 & +k & -\beta & +\beta \\ +\cos(ka) & +\sin(ka) & -e^{\beta b} & -e^{-\beta b} \\ +k\sin(ka) & -k\cos(ka) & +\beta e^{\beta b} & -\beta e^{-\beta b} \end{vmatrix} = 0 \quad (12)$$

In general, expanding a determinant with four rows and four columns will give 24 terms. However, for the determinant in Eq. 12, ten terms are zero to reduce this to 14 terms. Also setting the matrix elements that are plus or minus 1 in these 14 terms gives the simpler expression in Eq. 13 for the determinant.

$$Det = -M_{23}M_{32}M_{41} + M_{24}M_{32}M_{41} + M_{22}M_{33}M_{41} - M_{22}M_{34}M_{41} + M_{23}M_{31}M_{42}$$
$$- M_{24}M_{31}M_{42} - M_{24}M_{33}M_{42} + M_{23}M_{34}M_{42} - M_{22}M_{31}M_{43} + M_{24}M_{32}M_{43}$$
$$- M_{22}M_{34}M_{43} + M_{22}M_{31}M_{44} - M_{23}M_{32}M_{44} + M_{22}M_{33}M_{44} \quad (13)$$

Entering the expressions for each of the remaining terms in Eq. 13 gives Eq. 14.

$$Det = 4k\beta + \beta^2 \left(e^{\beta b} - e^{-\beta b}\right)\sin(ka)$$
$$-2k\beta \left(e^{\beta b} + e^{-\beta b}\right)\cos(ka) - k^2 \left(e^{\beta b} - e^{-\beta b}\right)\sin(ka) \quad (14)$$

Finally introducing hyperbolic functions, and the angle Θ defined in Eq. 15, we obtain Eq. 16 for the determinant which must be zero for a nontrivial solution.

$$\Theta \equiv ka \quad (15)$$

$$Det = 2\left(\beta^2 - k^2\right)\sin(\Theta)\sinh(\beta b) + 4k\beta\left[1 - \cos(\Theta)\cosh(\beta b)\right] \quad (16)$$

First, we consider two special cases for the determinant in Eq. 16.

Case 1: If Θ were any integer multiplied by 2π then Eq. 16 would simplify to give Eq. 17. Thus, the determinant may only be zero when βb or kβ is zero. Thus, b must be zero, or kβ is zero which would require that E is either zero or equal to $V_{II}$, which has no practical interest.

$$Det_1 = 4k\beta\left[1 - \cosh(\beta b)\right] \quad (17)$$

Case 2: If Θ were any odd integer multiplied by π then Eq. 16 would simplify to give Eq. 18. However, the hyperbolic cosine function has a minimum value of unity so the only solution would be for kβ to be zero. Thus, the only solution would be for E to be either zero or equal to $V_{II}$, which has no practical interest.

$$Det_2 = 4k\beta\left[1 + \cosh(\beta b)\right] \quad (18)$$

The following procedure is used for simulations with the square barrier model:

First specify the magnitude of the electrical charge for an electron and its mass and then specify the values of following three parameters for the model.
(1) the particle energy E.
(2) the potential $V_0$.

(3) the barrier length is which is the positive number b.
    Specify a trial value for the angle Θ, as a negative number in radians.
(1) Determine the sine and cosine of Θ.
(2) Use the value of E in Eq. 3 to determine k.
(3) Use the values of E and $V_0$ in Eq. 4 to determine β.
(4) Use the values of k and β in Eq. 16 to obtain the value of the determinant.
(5) Vary Θ to bring the determinant to zero to complete the solution.

Table I shows complete sets for the parameters (E, $V_0$, a, and b) from simulations using the procedure that was just defined for a closed circuit with a triangular barrier. At the zero crossing for each root there is a large jump between the positive and negative values for the determinant.

There is a continuous set of roots as a line between the values for the parameters in every pair of the adjacent rows. Larger steps in b were taken toward the base of the table to show that there are terminal values for b at both the top and the base of the table. Note that both ka, and thus Θ, are negative because a is defined to be a negative quantity as seen in Fig. 1.

Solutions with other values of E and/or $V_{II}$ would also form sets of non-intersecting line pairs so we anticipate that the complete solution would consist of an infinite number of non-intersecting surfaces in a bounded region of the parameter space. Thus, in may be best to make experiments near this boundary but within the parameter space the complete solution would be interpreted as being noise.

Table I. Complete solution for the parameters with a square
barrier where E is 0.95 eV and $V_{II}$ is 1.00 eV.

|  | Roots for Θ negative near 0 degrees | | | Roots for Θ negative near -360 degrees | | |
| --- | --- | --- | --- | --- | --- | --- |
| b, nm | Θ, deg | ka | a, nm | Θ, deg | ka | a, nm |
| 0.50 | -0.007529 | -0.000131 | -0.026320 | -359.8569 | -6.280688 | -1257.7377 |
| 1.00 | -0.015059 | -0.000263 | -0.052630 | -359.7189 | -6.278279 | -1257.2379 |
| 1.50 | -0.022588 | -0.000394 | -0.078950 | -359.5708 | -6.275694 | -1256.7382 |
| 2.00 | -0.030117 | -0.000526 | -0.105260 | -359.4278 | -6.273198 | -1256.3792 |
| 2.50 | -0.037647 | -0.000657 | -0.131578 | -359.2847 | -6.270701 | -1255.7380 |
| 3.00 | -0.045176 | -0.000788 | -0.157890 | -359.1417 | -6.268205 | -1255.1212 |
| 3.50 | -0.052706 | -0.000929 | -0.184210 | -358.9986 | -6.265708 | -1254.7379 |
| 4.00 | -0.060235 | -0.001051 | -0.210530 | -358.8560 | -6.263219 | -1254.2395 |
| 4.50 | -0.067764 | -0.001183 | -0.236840 | -358.7130 | -6.260723 | -1253.7362 |
| 5.00 | -0.075293 | -0.001314 | -0.263160 | -358.5695 | -6.258218 | -1253.2382 |
| 5.50 | -0.082822 | -0.001446 | -0.289470 | -358.4265 | -6.255722 | -1252.7383 |
| 6.00 | -0.090356 | -0.001577 | -0.315800 | -358.2834 | -6.253226 | -1252.2384 |
| 10.0 | -0.150585 | -0.002628 | -0.526310 | -357.1395 | -6.233260 | -1248.2402 |
| 20.0 | -0.301160 | -0.005256 | -1.052586 | -354.2827 | -6.183400 | -1238.2554 |
| 50.0 | -0.752720 | -0.013137 | -2.630836 | -345.7718 | -6,034856 | -1208.5089 |
| 100 | -1.504100 | -0.026251 | -5.256988 | -331.9912 | -5.794340 | -1160.3442 |
| 200 | -2.997900 | -0.052323 | -10.477976 | -307.1277 | -5.360390 | -1073.4437 |
| 500 | -7.320020 | -0.127758 | -25.584241 | -258.8950 | -4.519000 | -904.8653 |
| 1000 | -13.53910 | -0.236302 | -47.320581 | -227.8235 | -3.976268 | -796.2668 |
| 2000 | -21.21400 | -0.370254 | -74.145165 | -211.3954 | -3.689546 | -738.8492 |
| 5000 | -25.67990 | -0.448199 | -89.753955 | -206.0049 | -3.595464 | -720.0088 |

| 10000 | -25.84100 | -0.451011 | -90.317017 | -205.8425 | -3.592628 | -719.4409 |
| 20000 | -25.84190 | -0.451026 | -90.320163 | -205.8419 | -3.592619 | -719.4391 |
| 50000 | -25.84194 | -0.451027 | -90.320285 | -205.8419 | -3.592621 | -719.4395 |

### IIi. CLOSED-CIRCUIT WITH A TRIANGULAR BARRIER

Others have used Airy functions to solve the Schrödinger equation with one-dimensional unbounded models having a triangular potential barrier [1]. Now we use Airy functions with a one-dimensional model having finite linear extent as shown in Fig. 2 as a closed-circuit model. As in Fig. 1, the two ground symbols represent a connection with zero length to complete the circuit and the vertical line at the left end of the barrier may be thought of as a battery or other source of electrical potential.

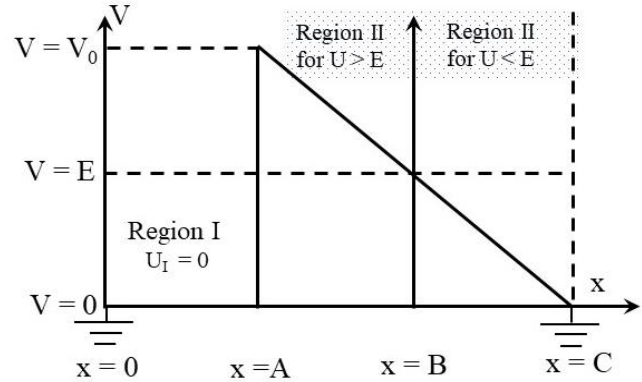

Fig. 2. Potential in regions I and II with a triangular barrier

To keep the derivations for the two models separate we use Eq. 17 for the wavefunction in Region I and again use the definition of k in Eq. 3.

$$\psi_I = C_1 \cos(kx) + C_2 \sin(kx) \qquad (17)$$

In Region II Airy functions [1] are required to solve the time-independent Schrödinger equation in Eq. 18 to determine the wave function. Here "e" is the magnitude of the particle charge so that E has units of electron volts and V has units of volts.

$$\frac{d^2\psi_{II}}{dx^2} + \frac{2me}{\hbar^2}[E - V(x)]\psi_{II} = 0 \qquad (18)$$

The electrical potential within the barrier is given by Eq. 19 as a linear interpolant so the x coordinate B at which the potential is equal to the energy of the particle is given by Eq. 20.

$$V = \frac{(C-x)V_0}{C-A} \qquad (19)$$

$$B = C - \frac{E}{V_0}(C-A) \qquad (20)$$

Substituting the potential from Eq. 19 into Eq. 18 gives Eq. 21, and the terms in Eq. 21 were regrouped to obtain Eq. 22.

$$\frac{d^2\psi_{II}}{dx^2} + \frac{2me}{\hbar^2}\left[E - \frac{V_0(C-x)}{(C-A)}\right]\psi_{II} = 0 \qquad (21)$$

$$\frac{d^2\psi_{II}}{dx^2} + \left[\frac{2me}{\hbar^2}\left[E - \frac{V_0 C}{(C-A)}\right] + \frac{2meV_0}{\hbar^2(C-A)}x\right]\psi_{II} = 0 \quad (22)$$

Next, we make a change of parameters in the solution for the wave function by Valée and Soares [2] shown as Eq. 23 where the argument is in Eq. 24. It is convenient to separate the argument into two parts as shown in Eq. 25 where the two coefficients are K and γ are defined in Eqs. 26 and 27.

$$\psi_{II}(x) = C_3 Ai(-\xi) + C_4 Bi(-\xi) \quad (23)$$

$$-\xi = \left(\frac{2meV_0}{\hbar^2(C-A)}\right)^{\frac{1}{3}}\left[C - (C-A)\frac{E}{V_0}\right] - \left(\frac{2meV_0}{\hbar^2(C-A)}\right)^{\frac{1}{3}} x \quad (24)$$

$$-\xi = K - \gamma x \quad (25)$$

$$K \equiv \left(\frac{2meV_0}{\hbar^2(C-A)}\right)^{\frac{1}{3}}\left[C - (C-A)\frac{E}{V_0}\right] \quad (26)$$

$$\gamma \equiv \left(\frac{2meV_0}{\hbar^2(C-A)}\right)^{\frac{1}{3}} \quad (27)$$

From Eq 17, in Region I the derivative of the wavefunction is given by Eq. 28.

$$\psi_I'(x) = -C_1 k \sin(kx) + C_2 k \cos(kx) \quad (28)$$

Thus, the wavefunctions and their derivatives just inside the two ends of Region I are given by Eqs. 29 to 32.

$$\psi_I(0^+) = C_1 \quad (29)$$
$$\psi_I'(0^+) = kC_2 \quad (30)$$
$$\psi_I(A^-) = \cos(kA)C_1 + \sin(kA)C_2 \quad (31)$$
$$\psi_I'(A^-) = -k\sin(kA)C_1 + k\cos(kA)C_2 \quad (32)$$

The wavefunctions and their derivatives just inside the two ends of Region II are given by Eqs. 33 to 36.

$$\psi_{II}(A^+) = Ai(K - \gamma A)C_3 + Bi(K - \gamma A)C_4 \quad (33)$$
$$\psi_{II}'(A^+) = -\gamma Ai'(K - \gamma A)C_3 - \gamma Bi'(K - \gamma A)C_4 \quad (34)$$
$$\psi_{II}(C^-) = Ai(K - \gamma C)C_3 + Bi(K - \gamma C)C_4 \quad (35)$$
$$\psi_{II}'(C^-) = -\gamma Ai'(K - \gamma C)C_3 - \gamma Bi'(K - \gamma C)C_4 \quad (36)$$

Next pairs of equations from the group of Eqs. 29 to 36 are used to form the following system of 4 simultaneous homogeneous equations in the 4 unknown coefficients.

From Eqs. 31 and 33, for $\psi_I(A^-)$ equal to $\psi_{II}(A^+)$

$$\cos(kA)C_1 + \sin(kA)C_2 - Ai(K - \gamma A)C_3 - Bi(K - \gamma A)C_4 = 0 \quad (37)$$

From Eqs. 32 and 34, for $\psi_I'(A^-)$ equal to $\psi_{II}'(A^+)$:

$$-k\sin(kA)C_1 + k\cos(kA)C_2 + \gamma Ai'(K - \gamma A)C_3 + \gamma Bi'(K - \gamma A)C_4 = 0 \quad (38)$$

From Eqs. 29 and 35, for $\psi_I(0^+)$ equal to $\psi_{II}(C^-)$:
$$C_1 - Ai(K - \gamma C)C_3 - Bi(K - \gamma C)C_4 = 0 \tag{39}$$
From Eqs. 30 and 36, for $\psi_I'(0^+)$ equal to $\psi_{II}'(C^-)$:
$$kC_2 + \gamma Ai'(K - \gamma C)C_3 + \gamma Bi'(K - \gamma C)C_4 = 0 \tag{40}$$
The dimensionless parameter R is defined in Eq. 41. We divide Eqs. 38 and 40 by k and use the definition of R to replace these two equations with Eqs. 42 and 43 which are dimensionless. Thus, Eqs. 37, 39, 42, and 43 are combined to form Eq. 44 as a dimensionless matrix equation.
$$R \equiv \frac{\gamma}{k} \tag{41}$$
$$-\sin(kA)C_1 + \cos(kA)C_2 + RAi'(K - \gamma A)C_3 + RBi'(K - \gamma A)C_4 = 0 \tag{42}$$
$$C_2 + RAi'(K - \gamma C)C_3 + RBi'(K - \gamma C)C_4 = 0 \tag{43}$$
$$\begin{bmatrix} \cos(kA) & \sin(kA) & -Ai(K - \gamma A) & -Bi(K - \gamma A) \\ -\sin(kA) & \cos(kA) & +RAi'(K - \gamma A) & +RBi'(K - \gamma A) \\ 1 & 0 & -Ai(K - \gamma C) & -Bi(K - \gamma C) \\ 0 & 1 & +RAi'(K - \gamma C) & +RBi'(K - \gamma C) \end{bmatrix} \begin{bmatrix} C_1 \\ C_2 \\ C_3 \\ C_4 \end{bmatrix} = \begin{bmatrix} 0 \\ 0 \\ 0 \\ 0 \end{bmatrix} \tag{44}$$
To simplify the notation, we define the three parameters $\Theta$, X, and Y, where $\Theta$ has units of radians and X and Y are dimensionless, in Eqs. 45, 46, and 47, to modify Eq. 44 to obtain Eq. 48.
$$\Theta \equiv kA \tag{45}$$
$$X \equiv K - \gamma A \tag{46}$$
$$Y \equiv K - \gamma C \tag{47}$$
$$\begin{bmatrix} \cos(\Theta) & \sin(\Theta) & -Ai(X) & -Bi(X) \\ -\sin(\Theta) & \cos(\Theta) & +RAi'(X) & +RBi'(X) \\ 1 & 0 & -Ai(Y) & -Bi(Y) \\ 0 & 1 & +RAi'(Y) & +RBi'(Y) \end{bmatrix} \begin{bmatrix} C_1 \\ C_2 \\ C_3 \\ C_4 \end{bmatrix} = \begin{bmatrix} 0 \\ 0 \\ 0 \\ 0 \end{bmatrix} \tag{48}$$
Equation 48 is homogeneous so the determinant of the matrix must be zero as shown in Eq. 49 to have a non-trivial solution for the coefficients. Thus, the solutions must be determined by varying the three parameters X, Y, and $\Theta$ to bring the determinant to zero.
$$\begin{vmatrix} \cos(\Theta) & \sin(\Theta) & -Ai(X) & -Bi(X) \\ -\sin(\Theta) & \cos(\Theta) & +RAi'(X) & +RBi'(X) \\ 1 & 0 & -Ai(Y) & -Bi(Y) \\ 0 & 1 & +RAi'(Y) & +RBi'(Y) \end{vmatrix} = 0 \tag{49}$$
From Eq. 25, at x equal to A the argument in the Airy functions is given by Eq. 50, and for x equal to C the argument is given by Eq. 51.
$$-\xi(A) = K - \gamma A \tag{50}$$
$$-\xi(C) = K_1 - \gamma C \tag{51}$$
Substituting these two expressions for minus $\xi$ into Eq. 23 gives Eqs. 52 for X and Eq. 53 for Y, where K and $\gamma$ were defined in Eqs. 26 and 27. By using these two definitions we obtain

Eqs. 54, which is simplified to give Eq. 55 for X, and Eq. 56 for Y. Note that X is greater than or equal to zero, whereas Y is less than or equal to zero. Also, X minus Y which is independent of the energy and the potential, is given in Eq. 57.

$$X = K - \gamma A \quad (52)$$

$$Y = K - \gamma C \quad (53)$$

$$X = \left(\frac{2meV_0}{\hbar^2(C-A)}\right)^{\frac{1}{3}} \left[C - (C-A)\frac{E}{V_0}\right] - \left(\frac{2meV_0}{\hbar^2(C-A)}\right)^{\frac{1}{3}} A \quad (54)$$

$$X = \left(\frac{2meV_0}{\hbar^2}\right)^{\frac{1}{3}} (C-A)^{\frac{2}{3}} \left(1 - \frac{E}{V_0}\right) \quad (55)$$

$$Y = -\left(\frac{2meV_0}{\hbar^2}\right)^{\frac{1}{3}} (C-A)^{\frac{2}{3}} \frac{E}{V_0} \quad (56)$$

$$X - Y = \left(\frac{2meV_0}{\hbar^2}\right)^{\frac{1}{3}} (C-A)^{\frac{2}{3}} \quad (57)$$

Next these Eqs. 3, 20, 26, 27, 41, 46, 47, and 49 are used to define a procedure providing a unique and complete solution when three independent parameters are specified. Then, this procedure is implemented in an example. This demonstrates that there is one and only one solution whenever the following three parameters are specified: the particle energy, the peak value of the barrier potential, and the length of the triangular barrier.

The following procedure is used for our simulations with the triangular barrier model. Note that this is essentially the same as the method that was used with the square barrier model, but adapted for use with the triangular barrier.

First specify the magnitude of the electrical charge for an electron and its mass and then specify the values of following three parameters for the model.
(1) the particle energy E.
(2) the peak potential $V_0$.
(3) the barrier length which is C minus A.

Specify a trial value for the angle Θ with units of radians.
(1) Determine the sine and cosine of Θ.
(2) Use E and $V_0$ in Eq. 3 to determine k.
(3) Divide Θ by k to determine A.
(4) Add A to C minus A to determine C.
(5) Use C, A, and $V_0$ in Eq. 20 to determine B.
(6) Use $V_0$, C, and A in Eq. 27 to determine γ.
(7) Use k and γ in Eq. 41 to determine R.
(8) Use C, A, E and $V_0$ in Eq. 26 to determine K.
(9) Use A, K and γ in Eq. 46 to determine X.
(10) Use C, K and γ in Eq. 47 to determine Y.
(11) Calculate the Airy functions Ai, Bi, Ai' and Bi' for both X and Y in Eq. 49.
(12) Use R from step 7, the sine and cosine of Θ from step 1, and the Airy functions from step 11 to evaluate the determinant using Eq. 49. Then repeat these 12 steps using other values of Θ to bring the determinant to zero to complete one solution.

This process may be repeated using other sets of the three parameters, E, $V_0$, and C minus A to complete a desired set of solutions. Notice that in this procedure we specify only the three parameters E, $V_0$, and C-A. Then we vary $\Theta$ to bring the determinant to zero, thus determining the parameters A, C, B, $\gamma$, R, K, X, and Y to complete the solution. Thus, each set of values that may be specified for the three parameters E, $V_0$, and C-A corresponds to a unique and complete set for all of the other parameters.

Table II shows complete sets for the parameters (E, $V_0$, A, and B) from simulations using the procedure that was just defined for a closed circuit with a triangular barrier. Note that the only row in which the determinant is zero is for $\Theta$ equal to 180 degrees. Extra rows are provided in the table for 179 and 181 degrees to provide greater resolution at that point in these data.

Table II. Complete solution for the parameters with a triangular barrier where E is 0.95 eV, $V_0$ is 1.00 eV, and C-A is 2 nm.

| $\Theta$, deg | A, nm | B, nm | C, nm | determinant |
|---|---|---|---|---|
| 10 | 34.9511 | 35.0511 | 36.9511 | $-2.92805 \times 10^{-2}$ |
| 20 | 69.9021 | 70.0021 | 71.9021 | $-5.76712 \times 10^{-2}$ |
| 30 | 104.853 | 104.953 | 106.853 | $-8.43097 \times 10^{-2}$ |
| 40 | 139.804 | 139.904 | 141.804 | $-1.08386 \times 10^{-1}$ |
| 50 | 174.755 | 174.855 | 176.755 | $-1.29170 \times 10^{-1}$ |
| 60 | 209.706 | 209.806 | 211.706 | $-1.46029 \times 10^{-1}$ |
| 70 | 244.657 | 244.757 | 246.657 | $-1.58450 \times 10^{-1}$ |
| 80 | 279.608 | 279.708 | 281.608 | $-1.66058 \times 10^{-1}$ |
| 90 | 314.559 | 314.659 | 316.559 | $-1.68619 \times 10^{-1}$ |
| 100 | 349.511 | 349.611 | 351.511 | $-1.66058 \times 10^{-1}$ |
| 110 | 384.462 | 384.562 | 386.462 | $-1.58450 \times 10^{-1}$ |
| 120 | 419.413 | 419.513 | 421.413 | $-1.46029 \times 10^{-1}$ |
| 130 | 454.364 | 454.464 | 456.364 | $-1.29170 \times 10^{-1}$ |
| 140 | 489.315 | 489.415 | 491.315 | $-1.08386 \times 10^{-1}$ |
| 150 | 524.266 | 524.366 | 526.266 | $-8.43097 \times 10^{-2}$ |
| 160 | 559.217 | 559.317 | 561.217 | $-5.76712 \times 10^{-2}$ |
| 170 | 594.168 | 594.268 | 596.168 | $-2.92805 \times 10^{-3}$ |
| 179 | 625.624 | 625.724 | 627.624 | $-2.94281 \times 10^{-3}$ |
| 180 | 629.119 | 629.219 | 631.119 | $-2.06584 \times 10^{-17}$ |
| 181 | 632.614 | 632.714 | 634.614 | $2.94281 \times 10^{-3}$ |
| 190 | 664.070 | 664.170 | 660.070 | $2.92805 \times 10^{-2}$ |
| 200 | 699.021 | 699.121 | 701.021 | $5.76712 \times 10^{-2}$ |
| 210 | 733.972 | 734.072 | 735.972 | $8.43097 \times 10^{-2}$ |
| 220 | 768.923 | 769.023 | 770.923 | $1.08386 \times 10^{-1}$ |
| 230 | 803.874 | 803.974 | 805.874 | $1.29170 \times 10^{-1}$ |
| 240 | 838.825 | 838.925 | 840.825 | $1.46029 \times 10^{-1}$ |
| 250 | 873.776 | 873.876 | 875.776 | $1.58450 \times 10^{-1}$ |
| 260 | 908.727 | 908.827 | 910.727 | $1.66058 \times 10^{-1}$ |
| 270 | 943.678 | 943.778 | 945.678 | $1.68619 \times 10^{-1}$ |
| 280 | 978.629 | 978.729 | 980.629 | $1.66058 \times 10^{-1}$ |
| 290 | 1013.58 | 1013.68 | 1015.58 | $1.58450 \times 10^{-1}$ |
| 300 | 1048.53 | 1048.63 | 1050.53 | $1.46029 \times 10^{-1}$ |
| 310 | 1083.48 | 1083.58 | 1085.48 | $1.29170 \times 10^{-1}$ |
| 320 | 1118.43 | 1118.53 | 1120.43 | $1.08386 \times 10^{-1}$ |
| 330 | 1153.38 | 1153.48 | 1155.38 | $8.43097 \times 10^{-2}$ |
| 340 | 1188.34 | 1188.44 | 1190.34 | $5.76712 \times 10^{-2}$ |
| 350 | 1223.29 | 1223.39 | 1225.29 | $2.92805 \times 10^{-2}$ |
| 360 | 1258.24 | 1258.34 | 1260.24 | $4.1368 \times 10^{-17}$ |

The periodicity of the sine and cosine functions for the determinant in Eq. 49 requires that there are roots at all integer multiples of 360 degrees. Furthermore, because all system parameters are now determined including B and A, Table II shows that the length for tunneling, which is B minus A, is 0.10 nm in this example.

## IV. SUGGESTIONS FOR INITIAL FABRICATION AND TESTING OF PROTOTYPES

We would suggest using a metal wire loop as the pre-barrier region with a gap as the tunneling junction. These loops could be made of metals that have extremely low scattering for electrons at nanoscale to enable coherent transport of the electrons instead of having a vacuum pre-barrier. Gall simulated the electron mean-free path $\lambda$ for 20 metallic elements having different bulk resistivities $\rho_0$ at room temperature and reported that $\lambda$ is greatest at 68.2 nm for beryllium and is 53.3 nm, 39.9 nm, and 37.7 nm, respectively, for silver, copper, and gold [3]. However, the apparent bulk electrical resistivity of these metals increases as the diameter is reduced because of increased scattering of the electrons at surfaces and grain boundaries [3],[4]. Thus, the size and shape should be chosen to mitigate this loss. This design would provide a triangular potential barrier to implement the models which require large pre-barrier lengths to have roots where the determinant is zero.

One possible configuration for a device using beryllium, silver, or the other metals we have listed is what we call the "Capital Omega Model" because of its shape. The circular loop described in the previous paragraph corresponds to the upper part of the symbol for Omega with the tunneling gap at its base. The two horizontal outward legs at the base of this symbol correspond to the two halves of a dipole antenna to couple to the incident laser radiation used to power this device.

## V. SUMMARY AND CONCLUSIONS

We simulated quantum tunneling in two types of closed circuits. The first circuit has a square potential barrier across a pre-barrier region. The second has a triangular potential barrier across a pre-barrier region. The four-by-four matrix equation for each of these two circuits is homogeneous in the four parameters so its determinant must be zero for non-trivial solutions. Thus, with each circuit any three parameters may be specified and the fourth is varied to bring the determinant to zero to obtain the solution.

Table I shows the results with the square barrier model where we specify E, $V_{II}$, and a. The values on each row have two sections for separate roots, where the roots to the right have much larger values of the pre-barrier length. These two solutions begin at the degenerate case where a is zero and $\Theta$ is either zero or $2\pi$. Since b is a continuous variable, both solutions are on lines through the data points which are shown.

Table II shows that the solutions with the triangular barrier are at the points where ka is equal to $n\pi$ where n is an integer. This is significantly different from the results with the first model having a square barrier. Applications using the triangular barrier would require large pre-barrier lengths or else relatively high energy electrons could be used with relatively high barriers.

We plan to define and test means for making measurements using parameters such as those shown in Table I and Table II. This will include extending our previous work on generating microwave harmonics by focusing a mode-locked laser on the tip-sample junction of an STM. Then, we generated microwave harmonics at integer multiples of the pulse repetition frequency (74.254 MHz) of a mode-locked laser focused on the tunneling junction of a scanning tunneling microscope (STM) at the Los Alamos National Laboratory [5]. A bias-T was used to measure these harmonics which are superimposed on the STM DC tunneling current. We measured up to the 200th harmonic which is at 14.85 GHz with a measured power of only $3.162 \times 10^{-18}$ W. The power at each harmonic rolls off as the inverse square of its frequency because of the measurement circuit. However, analysis suggests that within the tunneling junction of the STM the harmonics extend to terahertz frequencies [6],[7]. We have described measurement systems that would provide more efficient coupling in measuring the harmonics [8].

## VI. ACKNOWLEDGMENTS

We are grateful to the Center for Integrated Nanotechnologies (CINT) at the Los Alamos National Laboratory for enabling us to make measurements with a scanning tunneling microscope and femtosecond lasers. We are also grateful for funding by the National Science Foundation in their awards 0338928 (2004) and 0712564 (2007) which enabled us to make simulations that led to the recent effort presented in this paper.